\documentclass[12 pt]{article}
\usepackage[utf8x]{inputenc}
\usepackage{amsmath}
\usepackage{amsfonts}
\usepackage{amssymb}
\usepackage{empheq}
\usepackage[sort&compress,numbers]{natbib}
\usepackage{doi}
\hoffset= - 0.9in         

\newcommand{\f}{\frac}
\newcommand{\s}{\sqrt}

\newcommand{\e}{\epsilon}

\newcommand{\al}{\alpha}

\newcommand{\nn}{\nonumber}

\newcommand{\D}{\Delta}
\newcommand{\res}{\operatorname{Res}}

\usepackage{hyperref}

\hypersetup{
  colorlinks   = true, 
  urlcolor     = blue, 
  linkcolor    = blue, 
  citecolor   = red 
}
\usepackage{hyperref}
\usepackage{tikz}
\usepackage{epigraph}
\usepackage{float}

\begin{document}
\title{Analytic properties of the Virasoro modular kernel}
\author{{Nikita Nemkov}\thanks{{\small {\it Moscow Institute of Physics and Technology (MIPT), Dolgoprudny, Russia}
			and {\it Institute for Theoretical and Experimental Physics (ITEP), Moscow, Russia}; nnemkov@gmail.com} }
	\date{ }}
\date{\today}
\maketitle
\vspace{-5.0cm}

\begin{center}
	\hfill ITEP/TH-24/16\\
\end{center}

\vspace{3.5cm}
\begin{abstract}
On the space of generic conformal blocks the modular transformation of the underlying surface is realized as a linear integral transformation. We show that the analytic properties of conformal block implied by Zamolodchikov's formula are shared by the kernel of the modular transformation and illustrate this by explicit computation in the case of the one-point toric conformal block. 
\end{abstract}
\tableofcontents

\section{Introduction and results}
Conformal blocks (CBs) can be defined as universal parts of the holomorphically factorized CFT correlation functions \cite{BPZ}. They are recognized as a new independent class of special functions relevant for many problems in modern physics including gauge theories \cite{AGT}. This paper is concerned with properties of the toric Virasoro one-point conformal block which is hereafter referred to simply as conformal block. This special representative of CBs is in some sense the simplest one, although it captures a lot of the important properties present in its more sophisticated counterparts such as the spheric Virasoro blocks, $W_{N}$- and WZW-conformal blocks, superconformal blocks etc.  

Toric CB is naturally defined as the following trace
\begin{eqnarray}
B_{\D}(q)=\operatorname{Tr}_\D \left(q^{L_0-\f{c}{24}} V_{\D_e}\right) \label{CB trace def}
\end{eqnarray}
Here $q$ is the toric nome $q=e^{2\pi i \tau}$; $V_{\D_e}$ is the primary field of dimension $\D_e$, the external dimension; $\D$ is the internal dimension -- dimension of the Verma module over which the trace is taken; and finally $c$ is the central charge of the theory. We will suppress parameters $\D_e$ and $c$ in our notation. Definition \eqref{CB trace def} allows to compute CB as a series expansion in powers of $q$
\begin{eqnarray}
B_{\D}(q)=q^{\D-\f{c}{24}}\left(1+q\f{\D_e(\D_e-1)}{2\D}+O(q^2)\right) \label{CB exp}
\end{eqnarray}
In the present paper we only consider conformal blocks with \textit{generic} values of parameters.
Then, the $q$-expansion coefficients at arbitrary order are not known in a simple closed form\footnote{However, various complementary representations exist. As examples we mention the AGT-inspired representation via the Nerkasov functions \cite{AGT, Wyllard:2009hg, Mironov:2009by} and the expansion in terms of the global conformal blocks \cite{Perlmutter:2015iya} which are basically the hypergeometric functions.}. Nevertheless, some non-perturbative aspects of CB are developed. In particular, it is known that as a function of  the internal dimension $\D$ conformal block has only simple poles located at the Kac zeros $\D=\D_{r,s}$ \eqref{Kac zeros} and that the $q$-dependence of the corresponding residues is described by the CBs with specific external dimensions $\D=\D_{r,-s}$ (note that these are not the Kac zeros)
\begin{eqnarray}
\underset{\D=\D_{r,s}}{\res}B_{\D}(q)=R_{r,s}B_{\D_{r,-s}}(q) \label{CB residues}
\end{eqnarray}
where $R_{r,s}$ are certain explicit $q,\D$-independent multipliers \eqref{R initial def}. It is also possible to find the regular part of CB and extend \eqref{CB residues} to arbitrary $\D$ \cite{Zamolodchikov:Rec_a, Zamolodchikov:Rec_c, Fateev:2009aw, Poghossian:2009mk, Hadasz:2009db}
\begin{eqnarray}
B_{\D}(q)=\chi_\D(q)+\sum_{r,s\ge1}\f{R_{r,s}}{\D-\D_{r,s}}q^{\D-\D_{r,s}}B_{\D_{r,-s}}(q) \label{Zamolodchikov's formula}
\end{eqnarray}
where $\chi_\D(q)=q^{\D-\f{c-1}{24}}/\eta(q)$ is the Virasoro character\footnote{$\eta(q)$ is the Dedekind eta function $\eta(q)=q^{1/24}\prod_{n\ge1}(1-q^n)$.}. Interestingly, this equation provides a recurrent relation among $q$-expansion coefficients and can be used to compute CB order by order in $q$ without reference to the definition \eqref{CB trace def}.

Another non-perturbative property of CB is related to the modular transformations acting on the torus and generated by the $S:\tau\to -1/\tau$ and $T:\tau\to \tau+1$ moves. Invariance of the correlation functions together with the linear independence of CBs with different $\D$ ($B_\D(q)\sim q^\D$) imply that $S$ and $T$ are represented as linear integral transformations on the space of CBs. The $T$ transformation acts simply as a phase shift and will not be considered while the $S$ transformation is non-trivial. Denoting the kernel of the $S$-transformation by $M_{\D\D'}$ one writes
\begin{eqnarray}
B_{\D}(q)=\int _{\D'}M_{\D\D'}B_{\D'}(\widetilde{q}) \label{M def}
\end{eqnarray}
where $\widetilde{q}=q^{-2\pi i/\tau}$. Note that the lhs and the rhs in \eqref{M def} are defined as expansions in $q$ about \textit{different} points and hence one can not study this modular transformation perturbatively in $q$. Instead, the full $q$-dependence must be taken into account.

With these arrangements in place we can describe the main result of the present paper. We note that the analytic structure of conformal block \eqref{CB residues} implies \textit{the same} analytic structure for the modular kernel. Indeed, taking the residue of \eqref{M def} at $\D=\D_{r,s}$ and using \eqref{CB residues} we obtain
\begin{eqnarray}
R_{r,s}B_{\D_{r,-s}}(q) = \int_{\D'}\underset{\D=\D_{r,s}}{\res}M_{\D\D'}B_{\D'}(\widetilde{q})
\end{eqnarray}
In turn, conformal block $B_{\D_{r,-s}}(q)$ can itself be expanded via the modular transformed blocks
\begin{eqnarray}
B_{\D_{r,-s}}(q)=\int_{\D'}M_{\D_{r,-s}\D'}B_{\D'}(\widetilde{q})
\end{eqnarray}
Comparing the two above equations and making use of the linear independence of CBs with different $\D'$ one discovers that
\begin{eqnarray}
\boxed{\underset{\D=\D_{r,s}}{\res}M_{\D\D'}=R_{r,s}M_{\D_{r,-s}\D'}} \label{M residues}
\end{eqnarray} 
This equation represents a non-trivial constraint required by consistency of the CB analytic structure and modular properties. In the remainder of the text we explicitly check relation \eqref{M residues} to find complete agreement.
\section{Modular kernel}
\subsection{Notation}
We start by defining our notation. It is useful to introduce the Liouville-type variables $\al,\al', \mu, b$ replacing the original CFT data
\begin{gather}
c=1+6Q^2,\qquad Q=b+b^{-1},\nn\\
\D=Q^2/4-\al^2,\qquad \D'=Q^2/4-\al'^2,\qquad \D_e=\mu(Q-\mu)
\end{gather}
With a little abuse of notation we will use the same letters for functions of the original and the newly introduced variables. Note however that due to a non-trivial Jacobian of the transformation from $\D$ to $\al$ property \eqref{M residues} is slightly different in terms of $\al$, namely
\begin{eqnarray}
\underset{\al=\al_{r,s}}{\res} M_{\al\al'}=-\f{R_{r,s}}{2\al_{r,s}}M_{\al_{r,-s}\al'} \label{M res al}
\end{eqnarray}
The Kac zeros are described by
\begin{eqnarray}
\D_{r,s}=Q^2/4-\al^2_{r,s},\qquad \al_{r,s} = \f{rb+sb^{-1}}{2} \label{Kac zeros}
\end{eqnarray}
for $r,s\ge1$. We emphasize that there are no poles in CB at $\D_{r,-s}$ for $r,s\ge1$. Note however that $\D_{r,s}=\D_{-r,-s}$. Without loss of generality throughout this paper we assume that $r,s\ge1$.
The multipliers entering \eqref{CB residues} read
\begin{eqnarray}
R_{r,s}=A_{r,s}P_{r,s} \label{R initial def}
\end{eqnarray}
where
\begin{eqnarray}
A_{r,s}=\f12\underset{(n,m)\neq(0,0), (r,s)}{\prod_{n=1-r}^r\prod_{m=1-s}^s}\f1{nb+mb^{-1}},\qquad
P_{r,s}=\prod_{n=1-r}^r\prod_{m=1-s}^s\left(nb+mb^{-1}-\mu\right) \label{R def}
\end{eqnarray}

\subsection{Explicit formula}
Modular kernel for the toric Virasoro blocks is known in closed form as an integral \cite{PT3} or a series \cite{Nemkov:2015zha} representation. For our current purposes the most handy form is the following
\begin{eqnarray}
M_{\al\al'}=\f{V_{\al}}{V_{\al'}}n_{\al'}\mathcal{M}_{\al\al'} \label{M factor def}
\end{eqnarray}
where $V_\al$ is a convenient renormalization function\footnote{Double gamma $\Gamma_b$ and sine $S_b$ functions to be extensively used below are described in appendix \ref{special functions}.}
\begin{eqnarray}
V_{\al}=\f{\Gamma_b(Q+2\al)\Gamma_b(Q-2\al)}{\Gamma_b(\mu+2\al)\Gamma_b(\mu-2\al)} \label{V def}
\end{eqnarray}
$n_{\al'}$ is an $\al$-independent factor 
\begin{eqnarray}
n_{\al'}=e^{2\pi i \mu\al'}\sin{2\pi b\al'}\sin{2\pi b^{-1}\al'}/S_b(\mu)
\end{eqnarray}
irrelevant for property \eqref{M res al}, while $\mathcal{M}_{\al\al'}$ is an essential contribution
\begin{align}
&\mathcal{M}_{\al\al'}=K_{\al\al'}+K_{-\al,\al'} \label{M al def}\\ 
&K_{\al\al'}=e^{4\pi i\al\al'}\sum_{n,m\ge0}e^{8\pi i \al_{n,m}\al'}K_{nm}(\al),\quad K_{nm}(\al)=\f{S_b(2\al+2\al_{n,m}+\mu)S_b(2\al_{n,m}+\mu)}{S_b(2\al+2\al_{n+1,m+1})S'_b(2\al_{n+1,m+1})} \label{K def}
\end{align}
Here $S'_b(z)$ denotes the derivative of $S_b(z)$. Now everything is set up and we can proceed to proving \eqref{M res al}.
\section{Proof of the residue formula}
We will prove assertion \eqref{M res al} by an explicit computation which appears to be straightforward but tedious. We would like to outline the important steps beforehand. Definition \eqref{M factor def} represents modular kernel as the product of normalization factor $V_{\al}/V_{\al'}$ \eqref{V def} and non-trivial series $\mathcal{M}_{\al\al'}$ \eqref{M al def}. It turns out that the normalization factor $V_{\al}$ features poles exactly at the Kac zeros and furthermore it satisfies
\begin{eqnarray}
\underset{\al=\al_{r,s}}{\res}V_{\al}=-\f{R_{r,s}}{2\al_{r,s}}V_{\al_{r,-s}} \label{V residues}
\end{eqnarray}
In contrast, the remainder $\mathcal{M}_{\al\al'}$ appears to be regular at $\al=\al_{r,\pm s}$ and to satisfy
\begin{eqnarray}
\mathcal{M}_{\al_{r,s},\al'}=\mathcal{M}_{\al_{r,-s},\al'} \label{M values}
\end{eqnarray}
Reconciled, these properties lead to \eqref{M res al}. In the rest of this section we show that equations \eqref{V residues} and \eqref{M values} hold.
\subsection{Normalization factor}
Let us compute the residue of $V_{\al}$ at $\al=\al_{r,s}$ and the value at $\al=\al_{r,-s}$. For $r,s\ge1$ which we assume without loss of generality, there is a single singular multiplier in $V_{\al}$ at $\al=\al_{r,s}$ while at $\al=\al_{r,-s}$ everything is regular (for summary of analytic properties of $\Gamma_b(z)$ see appendix \ref{special functions}). Note also that $\al_{r,s}+\al_{n,m}=\al_{r+n,s+m}$ and $Q=2\al_{1,1}$. Therefore, one writes
\begin{eqnarray}
\underset{\al=\al_{r,s}}{\res}V_{\al}=\f{\Gamma_b(2\al_{r+1,s+1})\underset{\al=\al_{r,s}}{\res}\Gamma_b(Q-2\al)}{\Gamma_b(\mu+2\al_{r,s})\Gamma_b(\mu+2\al_{-r,-s})},\qquad V_{\al_{r,-s}}=\f{\Gamma_b(2\al_{r+1,1-s})\Gamma_b(2\al_{1-r,s+1})}{\Gamma_b(\mu+2\al_{r,-s})\Gamma_b(\mu+2\al_{-r,s})}
\end{eqnarray}
The ratio reads
\begin{eqnarray}
\f{\underset{\al=\al_{r,s}}{\res}V_{\al}}{V_{\al_{r,-s}}}=\underbrace{\f{\underset{\al=\al_{r,s}}{\res}\Gamma_b(Q-2\al)}{\Gamma_b(2\al_{r+1,1-s})}}_{R_1} \underbrace{\f{\Gamma_b(2\al_{r+1,s+1})}{\Gamma_b(2\al_{1-r,s+1})}}_{R_2}\underbrace{\f{\Gamma_b(\mu+2\al_{r,-s})}{\Gamma_b(\mu+2\al_{-r,-s})}}_{R_3}\underbrace{\f{\Gamma_b(\mu+2\al_{-r,s})}{\Gamma_b(\mu+2\al_{r,s})}}_{R_4}
\end{eqnarray}
Let us calculate the first factor
\begin{multline}
R_1=\f{\underset{\al=\al_{r,s}}{\res}\Gamma_b(Q-2\al)}{\Gamma_b(2\al_{r+1,1-s})}=\lim_{\e\to0}\f{-\e}{2}\f{\Gamma_b(\al_{1-r,1-s}+\e)}{\Gamma_b(2\al_{r+1,1-s}+\e)}=\lim_{\e\to0}\f{-\e}{2}\prod_{n=1-r}^r\f{\Gamma(nb^2+1-s+b\e)}{\s{2\pi}b^{nb^2+1/2-s+\e}}=\\\lim_{\e\to0}\f{-\e}{2}\f{\Gamma(1-s+b\e)}{\s{2\pi}b^{1/2-s}}\prod_{\substack{n=1-r\\n\neq0}}^r\f{\Gamma(nb^2+1-s+\e)}{\s{2\pi}b^{nb^2+1/2-s+\e}}=\f{(-1)^{s}}{2\s{2\pi}b^{3/2-s}(s-1)!}\prod_{\substack{n=1-r\\n\neq0}}^r\f{\Gamma(nb^2+1-s)}{\s{2\pi}b^{nb^2+1/2-s}}
\end{multline}
Here the difference equation on the double gamma function \eqref{double gamma diff} was used.

Computation of the second factor is more straightforward as there is no limiting procedure involved
\begin{eqnarray}
R_2=\f{\Gamma_b(2\al_{r+1,s+1})}{\Gamma_b(2\al_{1-r,s+1})}=\prod_{n=1-r}^r\f{\s{2\pi}b^{nb^2+1/2+s}}{\Gamma(nb^2+1+s)}
\end{eqnarray}
Multiplying $R_1$ by $R_2$ one obtains
\begin{multline}
R_1\cdot R_2=\f{(-1)^{s}b^{2s-1}}{2(s-1)!s!}\prod_{\substack{n=1-r\\n\neq0}}^rb^{2s}\f{\Gamma(nb^2+1-s)}{\Gamma(nb^2+1+s)}=\f{(-1)^{s}b^{2s-1}}{2(s-1)!s!}\prod_{\substack{n=1-r\\n\neq0}}^r\prod_{m=1-s}^s\f{1}{nb+mb^{-1}}=\\ \f{(-1)^{s}b^{2s-1}}{2(s-1)!s!}\prod_{\substack{m=1-s\\m\neq0}}^{s}mb^{-1}\underset{(n,m)\neq (0,0)}{\prod_{n=1-r}^r\prod_{m=1-s}^s}\f{1}{nb+mb^{-1}}=-\f12 \underset{(n,m)\neq (0,0)}{\prod_{n=1-r}^r\prod_{m=1-s}^s}\f{1}{nb+mb^{-1}}
\end{multline}
which is exactly equal to $-\f{A_{r,s}}{2\al_{r,s}}$ with $A_{r,s}$ defined in \eqref{R def}. In a very similar manner one shows that the product of the remaining factors $R_3\cdot R_4$ is equal to $P_{r,s}$ defined in \eqref{R def}. Therefore, we conclude that relation \eqref{V residues} is satisfied.
\subsection{Regular part}
In order to prove \eqref{M res al} it remains to show that $\mathcal{M}_{\al\al'}$ is regular at $\al=\al_{r,\pm s}$ and satisfies \eqref{M values}. A care must be taken here since function $K_{\al\al'}$ \eqref{K def} is singular at these points, but the sum of $K_{\al\al'}$ and $K_{-\al\al'}$ defining $\mathcal{M}_{\al\al'}$ turns out to be regular. 
\subsubsection{Expansion near $\al=\al_{r,s}$}
Let us expand $\mathcal{M}_{\al\al'}$ near $\al=\al_{r,s}$
\begin{multline}
\mathcal{M}_{\al_{r,s}+\f\e2,\al'}=e^{-4\pi i\al_{r,s}}\left(e^{8\pi i\al_{r,s}\al'}e^{2\pi i\e\al'}K_{\al_{r,s}+\f\e2,\al'}+e^{-2\pi i\e\al'}K_{-\al_{r,s}-\f\e2,\al'}\right)=e^{-4\pi i\al_{r,s}}\sum_{n,m}e^{8\pi i\al_{n,m}\al'}\mathcal{M}^{r,s}_{n,m}(\e)
\end{multline}
where we have denoted
\begin{eqnarray}
\mathcal{M}^{r,s}_{n,m}(\e)=e^{2\pi i\e\al'}K_{n-r,m-s}\left(\al_{r,s}+\f{\e}2\right)\delta_{n\ge r, m\ge s}+e^{-2\pi i\e\al'}K_{n,m}\left(\al_{-r,-s}-\f{\e}2\right) \label{M rs coeff def}
\end{eqnarray}
Consider 
\begin{eqnarray}
K_{n-r,m-s}\left(\al_{r,s}+\f{\e}{2}\right)=\f{S_b(2\al_{n,m}+\mu+\e)S_b(2\al_{n-r,m-s}+\mu)}{S_b(2\al_{n+1,m+1}+\e)S_b'(2\al_{n-r+1,m-s+1})}
\end{eqnarray}
Taking into account that function $S_b(2\al_{n+1,m+1}+\e)$ has simple zero at $\e=0$ (see appendix \ref{special functions}) one writes the following small $\e$ expansion
\begin{multline}
e^{2\pi i\e\al'}K_{n-r,m-s}\left(\al_{r,s}+\f{\e}{2}\right)=\f{S_b(2\al_{n,m}+\mu)S_b(2\al_{n-r,m-s}+\mu)}{S'_b(2\al_{n+1,m+1})S_b'(2\al_{n-r+1,m-s+1})}\times \\ \left(\f{1}{\e}+2\pi i\al'+\f{S'_b(2\al_{n,m}+\mu)}{S_b(2\al_{n,m}+\mu)}-\f12\f{S''_b(2\al_{n+1,m+1})}{S'_b(2\al_{n+1,m+1})}+O(\e)\right) \label{K1 exp}
\end{multline}
Now turn to 
\begin{eqnarray}
K_{n,m}\left(\al_{-r,-s}-\f{\e}{2}\right)=\f{S_b(2\al_{n-r,m-s}+\mu-\e)S_b(2\al_{n,m}+\mu)}{S_b(2\al_{n-r+1,m-s+1}-\e)S_b'(2\al_{n+1,m+1})}
\end{eqnarray}
This term has different expansions depending on the balance of indices. If $n\ge r, m\ge s$ we have an expansion similar to \eqref{K1 exp}
\begin{multline}
e^{-2\pi i\e\al'}K_{n,m}\left(\al_{-r,-s}-\f{\e}{2}\right)=\f{S_b(2\al_{n-r,m-s}+\mu)S_b(2\al_{n,m}+\mu)}{S'_b(2\al_{n-r+1,m-s+1})S_b'(2\al_{n+1,m+1})}\times\\\left(-\f{1}{\e}+2\pi i\al'+\f{S'(2\al_{n-r,m-s}+\mu)}{S_b(2\al_{n-r,m-s}+\mu)}-\f12\f{S''_b(2\al_{n-r+1,m-s+1})}{S'_b(2\al_{n-r+1,m-s+1})}+O(\e)\right) \label{K2 exp}
\end{multline}
And we see that the sum of \eqref{K1 exp} and \eqref{K2 exp} is indeed regular at $\e=0$ and given by
\begin{multline}
\mathcal{M}_{n,m}^{r,s}(0)=\f{S_b(2\al_{n-r,m-s}+\mu)S_b(2\al_{n,m}+\mu)}{S'_b(2\al_{n-r+1,m-s+1})S_b'(2\al_{n+1,m+1})}\times\Big(4\pi i\al'+\\
\f{S'_b(2\al_{n,m}+\mu)}{S_b(2\al_{n,m}+\mu)}+\f{S'_b(2\al_{n-r,m-s}+\mu)}{S_b(2\al_{n-r,m-s}+\mu)}-\f12\f{S''_b(2\al_{n+1,m+1})}{S'_b(2\al_{n+1,m+1})}-\f12\f{S''_b(2\al_{n-r+1,m-s+1})}{S'_b(2\al_{n-r+1,m-s+1})}\Big),\\\qquad n\ge r, m\ge s \label{M rs large}
\end{multline}
When $n<r$ and $m<s$ function $S^{-1}_b(2\al_{n-r+1,m-s+1}-\e)=O(\e)$ so the second term in \eqref{M rs coeff def} vanishes at $\e=0$ while the first terms is absent due to factor $\delta_{n\ge r, m\ge s}$ hence 
\begin{eqnarray}
\mathcal{M}_{n,m}^{r,s}(0)=0,\qquad n<r, m<s \label{M rs small}
\end{eqnarray}
Finally, when $n<r$ and $m\ge s$ or $n\ge r$ and $m<s$ function $S^{-1}_b(2\al_{n-r+1,m-s+1}-\e)$ is regular at $\e=0$ and we have 
\begin{eqnarray}
\mathcal{M}_{n,m}^{r,s}(0)=\f{S_b(2\al_{n-r,m-s}+\mu)S_b(2\al_{n,m}+\mu)}{S_b(2\al_{n-r+1,m-s+1})S_b'(2\al_{n+1,m+1})},\qquad n<r, m\ge s \quad\text{or}\quad n\ge r,m<s \label{M rs medium}
\end{eqnarray}
\subsubsection{Expansion near $\al=\al_{r,-s}$}
Now let us expand $\mathcal{M}_{\al\al'}$ near $\al=\al_{r,-s}$.
\begin{multline}
\mathcal{M}_{\al_{r,-s}+\f\e2,\al'}=e^{-4\pi i\al_{r,s}}\left(e^{2\pi i\e\al'}e^{8\pi i\al_{r,0}\al'}K_{\al_{r,s}+\f\e2,\al'}+e^{-2\pi i\e\al'}e^{8\pi i\al_{0,s}\al'}K_{\al_{-r,s}-\f\e2,\al'}\right)=\\e^{-4\pi i\al_{r,s}}\sum_{n,m}e^{8\pi i\al_{n,m}\al'}\mathcal{M}^{r,-s}_{n,m}(\e)
\end{multline}
where we have denoted
\begin{eqnarray}
\mathcal{M}^{r,-s}_{n,m}(\e)=e^{2\pi i\e\al'}K_{n-r,m}\left(\al_{r,-s}+\f{\e}2\right)\delta_{n\ge r}+e^{-2\pi i\e\al'}K_{n,m-s}\left(\al_{-r,s}-\f{\e}2\right)\delta_{m\ge s} \label{M r-s coeff def}
\end{eqnarray}
We emphasize that $\mathcal{M}^{r,-s}_{n,m}(\e)$ is not obtained from $\mathcal{M}^{r,s}_{n,m}(\e)$ \eqref{M rs coeff def} by flipping the sign of $s$. 

Proceeding in full analogy with the previous subsection one shows that $\mathcal{M}^{r,-s}_{n,m}(\e)$ is regular at $\e=0$ with different expansions depending on $n-r$ and $m-s$. For $n\ge r,m\ge s$ one obtains
\begin{multline}
\mathcal{M}^{r,-s}_{n,m}(0)=\f{S_b(2\al_{n,m-s+\mu})S_b(2\al_{n-r,m+\mu})}{S_b'(2\al_{n-r+1,m+1})S_b'(2\al_{n+1,m-s+1})}\times \Big(4\pi i\al'+\\\f{S_b'(2\al_{n-r,m}+\mu)}{S_b(2\al_{n-r,m}+\mu)}+\f{S_b'(2\al_{n,m-s}+\mu)}{S_b(2\al_{n,m-s}+\mu)}-\f12 \f{S_b''(2\al_{n+1,m-s+1})}{S_b'(2\al_{n+1,m-s+1})}-\f12 \f{S_b''(2\al_{n-r+1,m+1})}{S_b'(2\al_{n-r+1,m+1})}\Big),\\\quad n\ge r,m\ge s \label{M r-s large}
\end{multline}
When $n<r$ and $m<s$ $\mathcal{M}^{r,-s}_{n,m}(0)$ is vanishing due to the Kronecker deltas 
\begin{eqnarray}
\mathcal{M}_{n,m}^{r,-s}(0)=0,\qquad n<r, m<s \label{M r-s small}
\end{eqnarray}
and finally when $n<r$ and $m\ge s$ or $n\ge r$ and $m<s$ there are no singular terms and one has
\begin{eqnarray}
\mathcal{M}_{n,m}^{r,-s}(0)=\f{S_b(2\al_{n-r,m}+\mu)S_b(2\al_{n,m-s}+\mu)}{S_b(2\al_{n-r+1,m+1})S_b'(2\al_{n+1,m-s+1})},\qquad n<r, m\ge s \label{M r-s medium 1}\\
\mathcal{M}_{n,m}^{r,-s}(0)=\f{S_b(2\al_{n-r,m}+\mu)S_b(2\al_{n,m-s}+\mu)}{S_b(2\al_{n+1,m-s+1})S_b'(2\al_{n-r+1,m+1})},\qquad n\ge r,m<s \label{M r-s medium 2}
\end{eqnarray}
\subsubsection{Comparison}
Let us first compare equations \eqref{M rs large} and $\eqref{M r-s large}$ which are valid for $n\ge r, n\ge s$. Consider the ratio of $\mu$-dependent terms in the overall prefactors. Using property \eqref{double sine diff} one obtains
\begin{eqnarray}
\f{S_b(2\al_{n,m-s}+\mu)S_b(2\al_{n-r,m}+\mu)}{S_b(2\al_{n-r,m-s}+\mu)S_b(2\al_{n,m}+\mu)}=\f{\prod\limits_{k=n-r}^{n-1}2\sin{\pi(kb^2+\mu b+m-s)}}{\prod\limits_{k=n-r}^{n-1}2\sin{\pi(kb^2+\mu b+m)}}=(-1)^{rs}
\end{eqnarray}
Now, differentiating \eqref{double sine diff}, substituting $z={2\al_{n+1,m+1}}$, and taking into account that $S_b(2\al_{n+1,m+1})=0$ gives
\begin{eqnarray}
S'_b(2\al_{n+2,m+1})=2\sin 2\pi b\al_{n,m}S'_b(2\al_{n+1,m+1}) \label{double sine d diff}
\end{eqnarray}
Using this equation one computes the ratio of the remaining $\mu$-independent terms in the overall prefactors of \eqref{M rs large} and \eqref{M r-s large}
\begin{eqnarray}
\f{S'_b(2\al_{n-r+1,m-s+1})S'_b(2\al_{n+1,m+1})}{S_b'(2\al_{n+1,m-s+1})S_b'(2\al_{n-r+1,m+1})}=\f{\prod\limits_{k=n-r+1}^{n} 2\sin \pi(kb^2+m+1)}{\prod\limits_{k=n-r+1}^{n} 2\sin \pi(kb^2+m-s+1)}=(-1)^{rs}
\end{eqnarray}
Hence, the overall factors are the same in \eqref{M rs large} and \eqref{M r-s large} and the $\al'$-dependent terms agree exactly. Now, consider 
\begin{eqnarray}
T^{r,s}_{n,m}=\f{S'_b(2\al_{n-r,m-s}+\mu)}{S_b(2\al_{n-r,m-s}+\mu)}+\f{S'_b(2\al_{n,m}+\mu)}{S_b(2\al_{n,m}+\mu)}\\
T^{r,-s}_{n,m}=\f{S'_b(2\al_{n-r,m}+\mu)}{S_b(2\al_{n-r,m}+\mu)}+\f{S'_b(2\al_{n,m-s}+\mu)}{S_b(2\al_{n,m-s}+\mu)}
\end{eqnarray}
which enter expressions \eqref{M rs large} and \eqref{M r-s large} respectively. We will show by induction in $r,s$ that these functions coincide. For $r,s=0$ this is trivial. Assume that $T^{r,s}_{n,m}=T^{r,-s}_{n,m}$ for some $r, s$ and consider
\begin{multline}
T^{r+1,s}_{n,m}=\f{S'_b(2\al_{n-r+1,m-s}+\mu)}{S_b(2\al_{n-r-1,m-s}+\mu)}+\dots=\\\f{2\sin\pi b(2\al_{n-r,m-s}+\mu)}{S_b(2\al_{n-r,m-s}+\mu)}\times\\\f{S'_b(2\al_{n-r,m-s}+\mu)-2\pi b \cos{\pi b(2\al_{n-r-1,m-s}+\mu)}S_b(2\al_{n-r-1,m-s}+\mu)}{2\sin\pi b(2\al_{n-r,m-s}+\mu)}+\dots=\\T^{r,s}_{n,m}-\pi b\cot\pi ((n-r-1)b^2+\mu b)
\end{multline}
where in the intermediate steps the $r$-independent part of $T^{r,s}_{n,m}$ is denoted by ellipses. Also, besides relation \eqref{double sine diff} we have used relation
\begin{eqnarray}
S'_b(z+b)=2\sin \pi bz S_b'(z)+2\pi b\cos\pi b z S_b(z)
\end{eqnarray}
which is obtained by differentiating \eqref{double sine diff}. Mimicking the above computation one shows that
\begin{eqnarray}
T^{r+1,-s}_{n,m}=T^{r,-s}_{n,m}-\pi b\cot\pi ((n-r-1)b^2+\mu b)
\end{eqnarray}
And hence $T^{r+1,s}_{n,m}=T^{r+1,-s}_{n,m}$. Induction in $s$ proceeds in full analogy and we will omit it. 

A last step in verifying agreement between \eqref{M rs large} and \eqref{M r-s large} is to show that functions
\begin{eqnarray}
U^{r,s}_{n,m}=\f{S''_b(2\al_{n-r+1,m-s+1})}{S'_b(2\al_{n-r+1,m-s+1})}+\f{S''_b(2\al_{n+1,m+1})}{S'_b(2\al_{n+1,m+1})}\\
U^{r,-s}_{n,m}=\f{S''_b(2\al_{n-r+1,m+1})}{S'_b(2\al_{n-r+1,m+1})}+\f{S''_b(2\al_{n+1,m-s+1})}{S'_b(2\al_{n+1,m-s+1})}
\end{eqnarray}
also coincide. We confine ourselves to pointing out that the following property
\begin{eqnarray}
S_b''(2\al_{n+2,m+1})=2\sin 2\pi b\al_{n+1,m+1} S_b''(2\al_{n+1,m+1})+4\pi b\cos2\pi b\al_{n+1,m+1}S'_b(2\al_{n+1,m+1}) \label{double sine dd diff}
\end{eqnarray}
reduces comparison of $U^{r,s}_{n,m}$ and $U^{r,-s}_{n,m}$ to basically the same computation that we carried out for $T^{r,\pm s}_{n,m}$ and we will not present it here. Formula \eqref{double sine dd diff} is obtained from \eqref{double sine diff} by double differentiation and substitution of $z=2\al_{n+1,m+1}$ together with using $S_b(2\al_{n+1,m+1})=0$.

Hence we have shown that
\begin{eqnarray}
\mathcal{M}_{n,m}^{r,s}(0)=\mathcal{M}_{n,m}^{r,-s}(0),\qquad n\ge r, m\ge s
\end{eqnarray} 
Coincidence of these functions for $n<r,m<s$ is trivial since they both vanish \eqref{M rs small}, \eqref{M r-s small}. It remains to compare \eqref{M rs medium} against \eqref{M r-s medium 1}, \eqref{M r-s medium 2}. This is again a straightforward but somewhat bulky exercise making use of relations \eqref{double sine diff} and \eqref{double sine d diff}. We will omit the computation and only report a complete agreement.  This completes our proof of equation \eqref{M values}. 

\section{Discussion}
The non-perturbative aspects of CBs are hard to reveal. Zamolodchikov's formula \eqref{Zamolodchikov's formula} describing the analytic structure of CB to all orders in $q$ is a remarkable exception. The fact that the explicit expression for the modular kernel of generic CB is available \eqref{M factor def} is also quite non-trivial. It is instructive to recall how this expression is derived \cite{PT3, Nemkov:2015zha}. The algebra of modular transformations features non-linear consistency relations such as the pentagon and the hexagon identities and their toric counterparts \cite{MSlecturesRCFT}. Certain specifications of these non-linear relations give rise to linear difference equations on the generic modular kernel with \textit{degenerate} modular kernels entering as \textit{coefficients}. Degenerate CBs correspond to finite representations of the Virasoro algebra and satisfy the differential BPZ equations. They can be found exactly and the corresponding modular kernels (which are simply finite matrices) can be computed. Hence, in deriving formula \eqref{K def} only properties of a very special class of CBs was explicitly used, but the result is supposed to describe the modular transformations of generic CB. Validity of equation \eqref{M res al} following from the analytic structure of generic CB furnishes a highly non-trivial test of this assertion. 

Moreover, these equations partly explain an unexpected structure of expression \eqref{K def}. As confirmed from many perspectives \cite{GMMpert, Nemkov1, Lerda, Billo:2013fi, GMMnonpert, Nemkov2, Nemkov:2015zha} the Fourier-type contribution $e^{4\pi i\al\al'}$ is always present in the modular kernel at the perturbative level\footnote{Here the perturbative expansion in inverse powers of $\D$ as $\D\to\infty$ is implied. This should not be confused with the perturbative $q$-expansion which we usually discuss in the text.}. From this point of view, the expansion in \eqref{K def} looks like a non-perturbative completion with powers of parameters $e^{4\pi i b\al}, e^{4\pi i b^{-1}\al}$ which do not appear in $q$-expansion of CB. Equation \eqref{M res al}, valid on general grounds, can not be satisfied by the Fourier kernel alone and necessitates the introduction of the aforementioned non-perturbative terms.

One more remark is in order. Zamolodchikov's relation for CB is powerful enough to replace the definition and give an efficient computational approach. Although we already have an explicit formula for the modular kernel it is interesting to understand whether a relation similar to Zamolodchikov's recursion can be found for the modular kernel based solely on property \eqref{M res al}. We argue that this is not the case. It is the non-trivial series expansion part of the modular kernel $\mathcal{M}_{\al\al'}$ \eqref{M al def} for which we would like to obtain a recursive definition. However, this part is regular and only satisfies condition \eqref{M values}. This is not enough to construct a recurrence equation valid for all $\al$. In other words, in the full modular kernel $M_{\al\al'}$ \eqref{M factor def} all the poles come from an $\al'$-independent normalization factor $V_{\al}$, and hence they are common to all coefficients of the expansion that we wish to describe. In contrast, in the expansion of conformal block \eqref{CB exp} additional poles appear as the order of $q$ increases. This is the reason why \eqref{CB residues} relates different orders of the $q$-expansion allowing for recursive computations.

Finally, we would like to stress that although we have only checked formula \eqref{M residues} for the toric Virasoro block the derivation is very general and extensions to many other cases should exist. For example, the spheric modular kernel should satisfy \eqref{M residues} if the toric residue coefficients $R_{r,s}$ are replaced by their spheric counterparts.

\subsection*{Acknowledgements}
The author is grateful to Alexei Morozov and Andrey Mironov for their guidance. The work is partly supported by grants RFBR 16-01-00291, RFBR 16-32-00920-mol-a, RFBR 15-51-52031-NSC-a, RFBR 16-51-53034-GFEN, RFBR 15-51-50034-YaF, and MK-8769.2016.1.

\appendix
\section{Double gamma and sine functions \label{special functions}}
Double gamma function $\Gamma_b(z)$ can be defined as the analytic continuation of the following integral
\begin{eqnarray}
\Gamma_b(z)=\int_0^\infty \f{dt}{t}\left(\f{e^{-zt}-e^{-Qt/2}}{(1-e^{-bt})(1-e^{-b^{-1}t})}-\f{(Q-2z)^2}{8e^t}-\f{Q-2z}{2t}\right),\qquad Q=b+b^{-1} \label{double gamma def}
\end{eqnarray}
$\Gamma_b(z)$ is meromorphic with no zeros and only simple poles located at $z=-nb-mb^{-1}$ for $n,m\ge0$, i.e. schematically $\Gamma_b(z)\propto \prod_{n,m\ge0}\f1{z+rb+sb^{-1}}$. Double gamma function satisfies the following difference equations
\begin{align}
\Gamma_b(z+b)=\Gamma_b(z)\f{\s{2\pi}b^{bz-1/2}}{\Gamma(bz)},\qquad \Gamma_b(z+b^{-1})=\Gamma_b(z)\f{\s{2\pi}b^{1/2-b^{-1}z}}{\Gamma(b^{-1}z)} \label{double gamma diff}
\end{align}
related to each other by the replacement $b\to b^{-1}$ which is a symmetry of the double gamma function $\Gamma_b(z)=\Gamma_{b^{-1}}(z)$.

Double sine function $S_b(z)$ is defined as
\begin{eqnarray}
S_b(z)=\f{\Gamma_b(z)}{\Gamma_b(Q-z)} \label{double sine def}
\end{eqnarray}
It shares poles with the double gamma function but possesses additional zeros at $z=nb+mb^{-1}$ for $n,m \ge1$, i.e. schematically $S_b(z)\propto \prod_{n,m\ge0}\f{z-(n+1)b-(m+1)b^{-1}}{z+nb+mb}$. Double sine function satisfies the following difference equations
\begin{eqnarray}
S_b(z+b)=2\sin(\pi b z) S_b(z),\quad S_b(z+b^{-1})=2\sin(\pi b^{-1} z) S_b^{-1}(z) \label{double sine diff}
\end{eqnarray}
and the symmetry property $S_b(z)=S_{b^{-1}}(z)$.
\bibliographystyle{utcaps_edited}
\bibliography{bibfile,revtex-custom}
\end{document}